\documentstyle[rawfonts,nassau,pslatex]{article}  
\frompage{000} \topage{000}                                              %| 
\def\rightmark{Probing Hadronization with Strangeness} 
\def\leftmark{S.A. Bass et al.}
 
\title{Probing Hadronization with Strangeness} 
\authors{
{S.A. Bass$^{1,2}$, M. Bleicher$^3$, J. Aichelin$^3$, F. Becattini$^4$,
 A. Ker\"anen$^5$, F.M. Liu$^3$, K. Redlich$^{6,7}$, K. Werner$^3$
}\\[2.812mm]
{\normalsize
\hspace*{-8pt}$^1$ Department of Physics, Duke University\\ 
	Durham, NC 27708-0305, USA\\[0.2ex] 
\hspace*{-8pt}$^2$ RIKEN-BNL Research Center, Brookhaven National Laboratory\\
	Upton, NY 11973, USA \\[0.2ex]
\hspace*{-8pt}$^3$ SUBATECH,
Laboratoire de Physique Subatomique et des
Technologies Associ\'ees \\
University of Nantes - IN2P3/CNRS - Ecole des Mines de Nantes \\
4 rue Alfred Kastler, F-44072 Nantes, Cedex 03, France\\[0.2ex]
\hspace*{-8pt}$^4$ Universit\'a di Firenze and INFN Sezione di Firenze \\
Via G. Sansone 1, I-50019, Sesto F.no, Florence, Italy\\[0.2ex]
\hspace*{-8pt}$^5$ Department of Physics, Theoretical Physics\\ 
90014 University of Oulu, Finland \\[0.2ex]
\hspace*{-8pt}$^6$ Theory Division, CERN \\
CH-1211 Geneva 23, Switzerland \\[0.2ex]
\hspace*{-8pt}$^7$ Institute for Theoretical Physics, University of Wroclaw \\ 
	PL-50204  Wroclaw, Poland \\[0.2ex]
}}
 
\abstract{ The $\overline{\Omega}/\Omega$ ratio originating from string
decays is predicted to be larger than unity
in proton-proton interaction at SPS energies. The anti-omega
dominance increases with decreasing beam energy. This surprising
behavior is caused by the combinatorics of quark-antiquark
production in small and low-mass  strings. Since this behavior is
not found in a statistical description of hadron production in
proton-proton collisions, it may serve as a key observable to
probe the hadronization mechanism in such collisions.}

 
\begin{document}
 
\maketitle

\section{Introduction}\label{intro}
Hadron yields and their ratios stemming from the final state of
ultra-relativistic heavy-ion collisions have been extensively used
to explore the degree of chemical equilibration
\cite{qgpreviews,stoecker,stock86a,rafelski,rafelski2,cleymans,johanna,braun-munzinger,spieles97a,bleicher02}
and to search for evidence for exotic
states and phase transitions in such collisions
\cite{qgpreviews}. Under the assumption of thermal and chemical
equilibrium, fits with a statistical (thermal) model have been
used to extract bulk properties of hot and dense matter, e.g. the
temperature and chemical potential at which chemical freeze-out
occurs
\cite{rafelski2,cleymans,johanna,braun-munzinger,spieles97a}.

The application of a statistical model to elementary hadron-hadron reactions
was first proposed by Hagedorn \cite{hag1} in order to describe the
exponential shape of the $m_t$-spectra of produced particles in p+p 
collisions. Recent analyses \cite{becattini97a} on hadron yields in 
electron-positron and proton-proton interactions at several centre-of-mass 
energies have shown that particle abundances as well can be 
described by a statistical ensemble with maximized entropy. In fact, the 
abundancies are consistent with a model assuming the existence of equilibrated 
fireballs at a temperature $T \approx 160-170$~MeV.
These findings have given renewed 
rise to the interpretation that hadronization 
in elementary hadron-hadron collisions is a purely statistical process,
which is difficult to reconcile with the popular dynamical picture that hadron 
production in pp collisions is due to the decay of color 
flux tubes \cite{colflux}.

In this article we argue  that  the $\overline{\Omega}/\Omega
\equiv \Omega^+/\Omega^-$ ratio in elementary proton-proton
collisions is an unambiguous and sensitive probe to distinguish between
particle production via the breakup of a color flux tube from
statistical hadronization \cite{omega_prl}.

\section{(Anti-)baryon production in sting models}
Color flux tubes, called strings, connect two SU(3) color charges
[ $3$ ] and [ $\overline 3$ ] with a linear confining potential.
If the excitation energy of the string is high enough it is
allowed to decay via the Schwinger mechanism \cite{schwinger},
i.e. the rate of newly produced quarks is given by:
\begin{equation}
\frac {{\rm d}N_{\kappa}}{{\rm d}p_{\perp}}
\sim {\rm exp}\left[-\pi m_{\perp}^2/\kappa \right]
\end{equation}
where $\kappa$ is the string tension and $m_{\perp} =
\sqrt{p_{\perp}^2 +m^2}$ is the transverse mass of the produced
quark with mass $m$.

However, specific string models may differ in their
philosophy and the types of strings that are created:
\begin{itemize}
\item
In UrQMD\cite{urqmdmodel} the projectile and target protons become
excited objects due to the momentum transfer in the interaction.
The resulting strings, with at most two strings being formed,  are
of diquark-quark type.
\item
In NeXuS\cite{nexusmodel}, the pp interaction is described in
terms of pomeron exchanges or ladder diagramms. Both hard and soft
interactions happen in parallel. Energy is shared equally between
all cut pomerons and the remnants. The endpoints of the cut
pomerons (i.e. the endpoints of the strings) may be valence
quarks, sea quarks, antiquarks or gluons. 
%The head partons emit
%additional partons due to the soft and hard evolution, thus
%getting harder and harder, until they meet in the center of the
%ladder. The emitted time-like partons then hadronize to final
%state paticles as do the hard partons in the center of the ladder.
\item
In PYTHIA\cite{pythiamodel}, a similar scheme as in UrQMD is employed. However,
hard interactions may create additional strings from scattered sea quarks.
Most strings are also of diquark-quark form.
\end{itemize}

\begin{figure}[h]
\vspace*{-0.8cm}
\insertplot{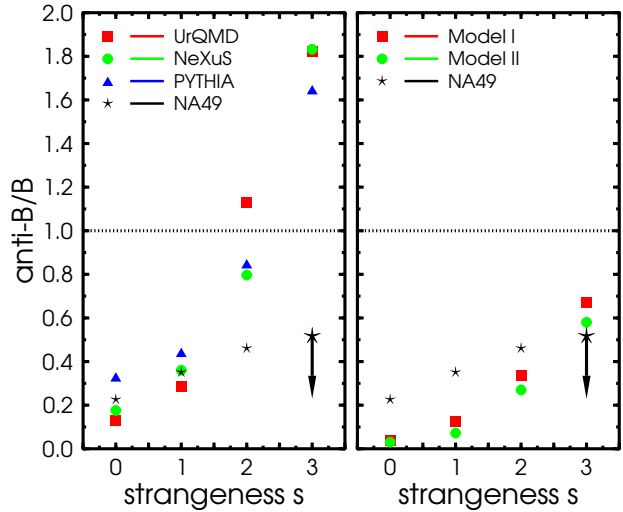}
\vspace*{-1.1cm}
\caption{Left: anti-baryon to baryon ratio {\it at midrapidity} in
pp interactions at 160 GeV as given by PYTHIA, NeXuS and UrQMD as
well as statistical model. Right: anti-baryon to baryon ratio in
$4\pi$ for the same reaction as given by statistical models. Stars
depict preliminary NA49 data for the $\overline B /B$ ratio at
midrapidity. \label{pythia}}
%\vspace{-0.4cm}
\end{figure}

Fig.\ref{pythia}a
 depicts the anti-baryon to baryon ratio at
midrapidity in proton-proton interactions at 160 GeV. The results
of the   calculations by NeXuS, UrQMD and PYTHIA, which are  the
best established string-fragmentation models for elementary
hadron-hadron interactions, are included in this figure. In all
these models, the $\overline B /B$ ratio increases strongly with
the strangeness content of the baryon. For strangeness $|s|=3$ the
ratio significantly exceeds unity. In UrQMD and PYTHIA the
hadronization of the diquark-quark strings leads directly to the
overpopulation of $\overline \Omega$. In NeXuS however, the
imbalance of quarks and anti-quarks in the initial state leads to
the formation of $q_{\rm val}-\overline s_{\rm sea}$ strings, (the
$s_{\rm val}-\overline q_{\rm sea}$ string is not possible). These
strings than result in the overpopulation of $\overline
\Omega$$^,$s. 

%............................................
%................................................

\section{(Anti-)baryon production in statistical models}
In Fig.\ref{pythia} the string model results are compared with the
predictions of statistical models (SM). Within the SM, hadron
productions is commonly described using the grand canonical (GC)
partition function, where the charge conservation is controlled by
the related chemical potential. In this description a net value of
a given U(1) charge is conserved on average. However, in the
limit of small particle multiplicities, conservation laws must
be implemented exactly, i.e., the canonical (C)
ensemble for conservation laws must be used
\cite{becattini97a,ko1,ahmed}. The conservation of quantum
numbers in the canonical approach severely reduces thermal  phase
space available for particle production. Thus, exact charge
conservation is of crucial importance in the description of
particle yields in proton induced processes and  in $e^+e^-$
\cite{becattini97a}, as well as in peripheral heavy ion collisions
\cite{ahmed}.

In Fig.\ref{pythia}   the predictions of two different canonical
statistical models for $\bar B/B$ ratios in pp collisions are
included. The main difference between these models is contained in
the implementation of baryon number and isospin conservation as
well as how additional strangeness suppression is
introduced.

({\bf I}) The calculation in this statistical model \cite{becalast}
is a full canonical one with fixed baryon number,  
strangeness and electric charge identical to those of initial state.  
Also, an extra strangeness suppression is needed to reproduce the  
experimental multiplicities. This is done by considering the number  
of newly produced $\langle s \overline s \rangle$ pairs as an additional  
charge to be found in the final hadrons. The $s\overline s$ pairs fluctuate  
according to a Poisson distribution and its mean number is considered 
a free parameter to be fitted \cite{becalast}. The parameters used  
for the prediction of $\Omega^+/\Omega^-$ ratio ($T$, global volume $V$ 
sum of single cluster volumes and $\langle s \overline s \rangle $)  
have been obtained by a fit to preliminary NA49 pp data  
\cite{na49pp} yielding $T=183.7 \pm 6.7$ MeV, $VT^3 = 6.49 \pm 1.33$  
and $\langle s \overline s \rangle = 0.405 \pm 0.026$ with a $\chi^2/dof= 
11.7/9$. It must be pointed out that $\Omega^+/\Omega^-$ ratio is  
actually independent of the $\langle s \overline s \rangle$ parameter 
and only depends on $T$ and $V$ (see also Fig.\ref{om}).

({\bf II}) Here we  first approximate the conservation of baryon
number and electric charge by the GC ensemble. Under  thermal
conditions at top SPS energies this approximation leads to deviations from
the exact C results in pp collisions  by at most 20-30$\%$
\cite{marek}.
Strangeness conservation  is, however,  implemented on the
canonical level following the procedure proposed in \cite{ahmed}.
It accounts for strong correlations of produced strange particles
due to constraints imposed by the locality of the conservation
laws. In pp collisions strangeness is not distributed in the whole
volume of the fireball  but is strongly correlated. A correlation
volume parameter $V_0=4\pi R_0^3/3$ is introduced, where $R_0\sim
1$ fm is a typical scale of QCD interactions.
Previous analysis of WA97 pA data yields: $R_0\sim 1.12$~fm
corresponding to $V_0\simeq 5.8$ fm$^3$. 
Note that hidden strange particles are not canonically
suppressed in this approach.
Analysis of experimental data in AA collisions has
shown that  $T$ and $\mu_B$ are almost entirely determined by the
collision energy and only dependent weakly on the number of
participants \cite{cleymans}. 
$4\pi$ results of
NA49 on $\bar p/\pi$ and $\pi /A_{part}$ ratios in pp and PbPb
collisions coincides within 20-30$\%$. In terms of the SM this
can be understood if $T$ and $\mu_B$ in pp and PbPb collisions have
similar values. 
We take $T\simeq 158$ MeV and $\mu_B\simeq 238$
MeV as obtained from SM analysis of full phase-space NA49 Pb-Pb
data \cite{becattini97a}. The volume of the fireball
 $V\sim 17$ fm$^3$ and the
charge chemical potential in pp was then found to reproduce the
average charge and baryon number in the initial state.

The predictions of the statistical models are shown in 
fig~\ref{pythia}(right).  
In these approaches the $\overline B /B$ ratio increases linearly with the  
strangeness content of the baryon. For comparison, both figures include  
preliminary data on the $\overline B /B$ ratio at midrapidity by NA49  
\cite{na49pp} (stars). 
Roughly 45 $\Omega^-$  have been so far extracted, 
no $\bar \Omega$ has been observed.
With a  95\% confidence level, the $\overline{\Omega}/\Omega$ 
ratio  in this measurement is below 0.5 \cite{NA49prelim}.
However, it must be stressed that in the statistical  
models one can calculate particle production only in full phase space and 
all quoted predictions refer to fully integrated particle spectra.

Large deviations of statistical models from the  data seen in
Fig.1b are to be expected as at midrapidity $\overline B /B$ ratios
are known to be much larger than in the whole phase space.  To
make predictions for particle ratios at midrapidity one can use
the canonical model {\bf II}. Following the procedure described in
\cite{ahmed} we first choose $T\simeq 168$ MeV, from the fit to
the midrapidity WA97 PbPb data \cite{johanna},  and $\mu_B\simeq
130$ MeV to reproduce $\bar p/p\simeq 0.22$ in pp collisions. With
these parameters the agreement of the  model {\bf II} with NA49
data is seen in Fig.1
to be quite satisfactory.

%............................................
\begin{figure}[t]
\vspace*{-0.5cm}
\insertplot{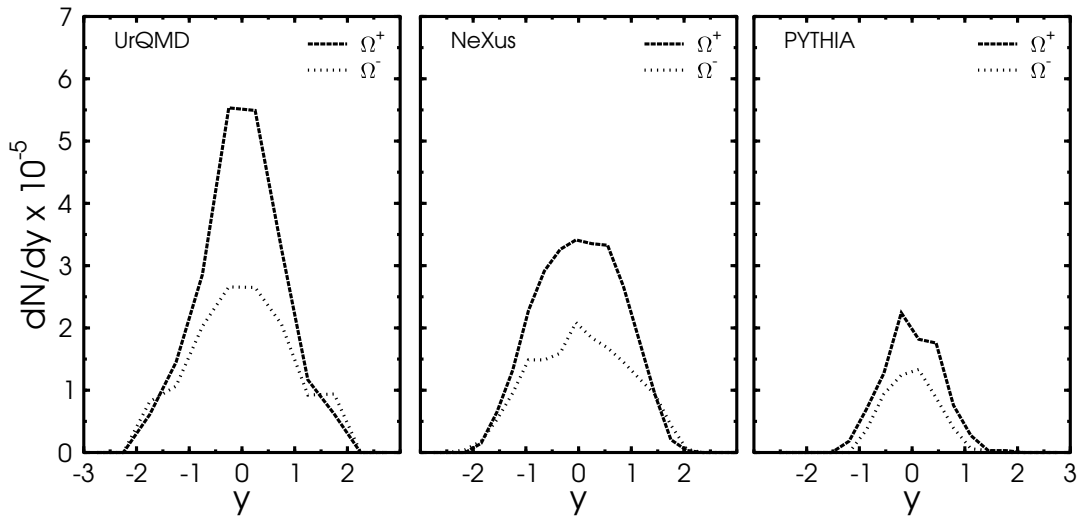} 
\vspace*{-1.1cm}
\caption{Rapidity density of anti-Omegas and Omegas in pp
interactions at 160 GeV as predicted by UrQMD, NeXus, PYTHIA.
\label{dndy}}
\vspace{-0.4cm}
\end{figure}
%................................................

\section{Rapidity-, mass- and volume dependencies}
The rapidity dependence of $\Omega$ and $\overline{\Omega}$ yield
is studied in Fig.~\ref{dndy} within different string models. The
results were calculated in pp interactions at 160 GeV within
PYTHIA, NeXuS and UrQMD (from top to bottom).
As can be seen, the $\overline{\Omega}/\Omega$ ratio is largest
around mid-rapidity.

The $\overline{\Omega}/\Omega$ ratio is fairly robust -- different
string-model implementations (PYTHIA, UrQMD, NeXuS) all agree in
their predictions within $\pm 20\%$. The value of
$\overline{\Omega}/\Omega
> 1$ in p+p reactions at the SPS is a generic feature of the
string-fragmentation. However, as shown in Table I the total
$4\pi$-yields of $\Omega$$^,$s and $\overline \Omega$$^,$s may
differ by a factor of four among the different string models. The
statistical models are in general giving more consistent results,
however deviations up to $20\%$   are not excluded.

%............................................
\begin{figure}[h]
\vspace*{-0.2cm}
\insertplot{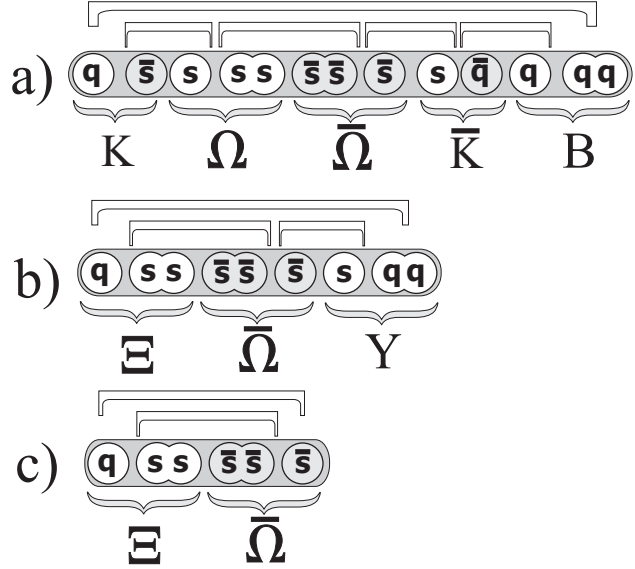} 
\vspace*{-0.5cm}
\caption{Fragmentation of a color field into quarks and hadrons.
While in large strings $\Omega$s and $\overline \Omega$s are
produced in equal abundance (top), small strings suppress
$\Omega$s at the string ends. \label{string}}
\vspace{-0.4cm}
\end{figure}
%................................................

In string models, the particle abundances depend on the parameters
chosen for the fragmentation scheme, while in statistical models
they reflect the differences between the ensembles chosen. Thus,
the absolute yields allow to distinguish between the
implementations once experimental data becomes available.

In order to   understand  the large  $\overline{\Omega}/\Omega$
values predicted by string models one elucidates  in
Fig.\ref{string}
the color flux tube break-up mechanism.  Fig.\ref{string} shows
the fragmentation of the color field into quark-antiquark pairs,
which then coalesce into hadrons. While in large strings
$\Omega$$^,$s and $\overline \Omega$$^,$s are produced in equal
abundance (a), low-mass strings in UrQMD suppress $\Omega$ production at
the string ends (b), while in NeXuS $\overline \Omega$$^,$s are enhanced (c).  
Thus, the microscopic method of hadronization
leads to a strong imbalance in $\overline{\Omega}/\Omega$ ratio in
low-mass strings.

\begin{figure}[h]
\vspace*{-0.5cm}
\insertplot{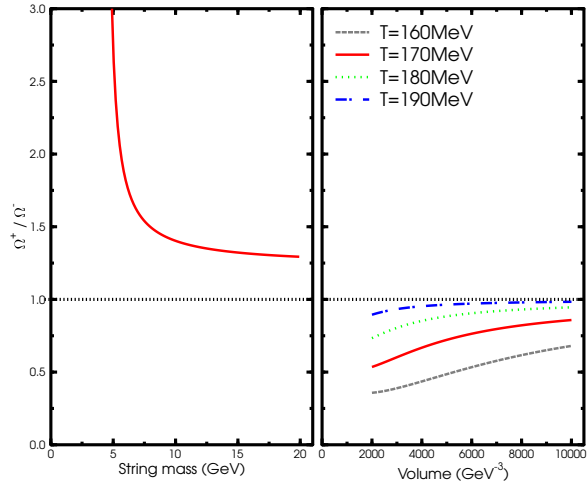}
\vspace*{-1.4cm}
\caption{Left: $\overline \Omega/\Omega$ ratio as a function of string mass.
Right: $\overline \Omega/\Omega$ ratio as a function of the 
volume in Modell II.
\label{om}}
\end{figure}

The $\overline{\Omega}/\Omega$ ratio depends in a strongly
non-linear fashion on the mass of the fragmenting string.
Fig.\ref{om} shows the $\overline \Omega/\Omega$ ratio as a
function of the mass of the fragmenting string (i.e. different
beam energies in pp). One clearly observes a strong enhancement of
$\bar{\Omega}$ production at low energies, while for large string
masses the ratio approaches the value  of
$\overline{\Omega}/\Omega = 1$ (which should be reached in the
limit of an infinitely long color flux tube).

However, it should be noted that recently a new class of string models
utilizing parton-based Gribov-Regge theory has been proposed which
are capable of generating an $\overline B /B$ ratio of less than
one \cite{fliu02}.

\begin{table}
\begin{tabular}{llrr}
&Model    & $\Omega$ ($\times 10^{-4}$)& $\overline
\Omega$ ($\times 10^{-4}$)\\\hline
&NeXus    & 0.48 & 0.79\\
&PYTHIA   & 0.17 & 0.30\\
&UrQMD    & 0.66 & 1.05\\
&Canonical Model {\bf I} & 0.46 & 0.31\\
&Canonical Model {\bf II}  &  0.41   &0.24 \\
\end{tabular}
\caption{\label{table1}$4\pi$ particles yields in pp collisions at 160GeV}
\end{table}

%............................................
%................................................

Statistical models, on the other hand, are not able to yield a ratio of  
$\overline{\Omega}/\Omega > 1$. This can be easily understood in the GC  
formalism, where $\overline{B}/B$ ratio is very sensitive to the baryon-chemical  
potential $\mu_B$. For finite baryon-densities, the $\overline{B}/B$ ratio 
will always be $< 1$ and only in the limit of $\mu_B = 0$ may  
$\overline{\Omega}/\Omega = 1$ be approached. These features survive in 
the canonical framework, where the GC fugacities are replaced by ratios of  
partition functions \cite{becattini97a}. This is shown in Fig.\ref{om}(right) 
where the ratio  
$\overline{\Omega}/\Omega$ in pp collisions (according to the previously  
described model {\bf I}) is plotted as a function of volume for four 
different temperatures. Hence, finite size corrections in the statistical  
model actually lead to the opposite behavior\cite{antik} in the ratio of  
$\overline{\Omega}/\Omega$ vs. system-size (i.e. volume replacing string-mass)  
than that observed in the fragmenting color flux tube picture.

\section{Balance Functions}
While the $\overline{\Omega}/\Omega$ ratio provides us with important 
information on the dynamics of hadronization, it does not yield any
information on the time-scales at which hadronization occurs: 
Balance functions offer a unique
model-independent formalism to probe the time-scales of a deconfined phase
and subsequent hadronization \cite{balance}.
Late-stage production of quarks could be attributed to three mechanisms:
formation of hadrons from gluons, conversion of the non-perturbative vacuum
energy into particles, or hadronization of a quark gas at constant
temperature. Hadronization of a quark gas should approximately conserve the net
number of particles due to the constraint of entropy conservation. Since
hadrons are formed of two or more quarks, creation of quark-antiquark pairs
should accompany hadronization. All three mechanisms for late-stage quark
production involve a change in the degrees of freedom. Therefore, any signal
that pinpoints the time where quarks first appear in a collision would provide
valuable insight into understanding whether a novel state of matter has been
formed and persisted for a substantial time.

The link between balance functions and the time at which quarks are created has
a simple physical explanation. Charge-anticharge pairs are created at the same
location in space-time, and are correlated in rapidity due to the strong
collective expansion inherent to a relativistic heavy ion collision. Pairs
created earlier can separate further in rapidity due to the higher initial
temperature and due to the diffusive interactions with other particles. The
balance function, which describes the momentum of the accompanying
antiparticle, quantifies this correlation. 
The balance function describes the conditional
probability that a particle in the bin $p_1$ will be accompanied by a particle
of opposite charge in the bin $p_2$:
%\vspace{-3mm}
\begin{equation}
\label{balancedef_eq}
B(p_2|p_1) \equiv \frac{1}{2}\left\{
\rho(b,p_2|a,p_1)-\rho(b,p_2|b,p_1) +\rho(a,p_2|b,p_1)-\rho(a,p_2|a,p_1)
\right\},
\end{equation}
%\vspace{-3mm}
where $\rho(b,p_2|a,p_1)$ is the conditional probability of observing a
particle of type $b$ in bin $p_2$ given the existence of a particle of type $a$
in bin $p_1$.  The label $a$ might refer to all negative kaons with $b$
referring to all positive kaons, or $a$ might refer to all hadrons with a
strange quark while $b$ refers to all hadrons with an antistrange quark. 
Balance functions will be discussed in greater detail in other 
articles of these proceedings \cite{scott_n,gary_n,marguerite_n}.

\section{Conclusions}\label{concl}
The $\overline \Omega/\Omega$ has been found to be extremely sensitive on
the dynamics of hadronization in p+p reactions.
Within the fragmenting color flux tube models we
have predicted that $\overline \Omega/\Omega$ ratio is
significantly above unity. This is in strong contrast to
statistical model results which  always imply   that $\overline
B / B $ ratios are bellow or equal to unity in proton+proton
reactions. Since this  observable is accessible by NA49
measurements at the SPS it can provide an excellent test to
distinguish the statistical model hadronization scenario from that
of microscopic color-flux tube dynamics -- first very preliminary
results seem to support the statistical hadronization hypothesis.
Balance functions yield complimentary information and 
offer a unique, model-independent formalism to probe
the time-scale of hadronization in heavy nucleus-nucleus collisions.

\section*{Acknowledgement(s)}
We thank K. Kadjia (NA49) for fruitful and stimulating
discussions. (S.A.B.) acknowledges support from RIKEN, Brookhaven
National Laboratory and DOE grants DE-FG02-96ER40945 as well as
DE-AC02-98CH10886. (K.R) acknowledges KBN grant, KBN-2P03B 03018.

\vfill\eject
\end{document}